\documentclass[conference]{IEEEtran}

\IEEEoverridecommandlockouts
\usepackage{cite}
\usepackage{amsmath,amssymb,amsfonts}
\usepackage{algorithmic}
\usepackage{graphicx}
\usepackage{textcomp}
\usepackage{xcolor}
\usepackage{hyperref}
\def\BibTeX{{\rm B\kern-.05em{\sc i\kern-.025em b}\kern-.08em
    T\kern-.1667em\lower.7ex\hbox{E}\kern-.125emX}}

\begin{document}
\pagestyle{empty}
\title{SDEX: Monte Carlo Simulation of Stochastic Differential Equations on Memristor Crossbars}
\author{Louis Primeau$^1$, Amirali Amirsoleimani$^2$, and Roman Genov$^1$ \\
$^1$Department of Electrical and Computer Engineering, University of Toronto, Toronto, Canada\\
$^2$Department of Electrical Engineering and Computer Science, York University, Toronto, Canada\\
Email: louis.primeau@mail.utoronto.ca, amirsol@yorku.ca, roman@eecg.utoronto.ca
}% <-this % stops a space
%\thanks{Manuscript received XX, XXXX; revised XX, XXXX.}

\markboth{ISCAS 2022,~Vol.XX, No.XX, May~2022}
{Shell \MakeLowercase{\textit{et al.}}: Bare Demo of IEEEtran.cls for IEEE Journals}

\maketitle
\thispagestyle{empty}
% As a general rule, do not put math, special symbols or citations
% in the abstract
\begin{abstract}
Here we present stochastic differential equations (SDEs) on a memristor crossbar, where the source of gaussian noise is derived from the random conductance due to ion drift in the memristor during programming. We examine the effects of line resistance on the generation of normal random vectors, showing the skew and kurtosis are within acceptable bounds. We then show the implementation of a stochastic differential equation solver for the Black-Scholes SDE, and compare the distribution with the analytic solution. We determine that the random number generation works as intended, and calculate the energy cost of the simulation.
 
\end{abstract}
%\vspace{-0.3cm}
% no keywords
\vspace{-0.3cm}
\IEEEpeerreviewmaketitle

\section{Introduction}

\IEEEPARstart{M}{emristor} crossbar arrays are an emerging CMOS alternative for acceleration of problems requiring compute-intensive matrix multiplication. Vector-matrix multiplication (VMM) operations can be done in
$\mathcal{O}(1)$ time complexity using Ohm's law to sum analog currents. Despite
their promise as VMM accelerators, current devices suffer from nonidealities stemming from
immature fabrication techniques \cite{design}, preventing perfectly accurate VMM
operations. Because of this, several previous works have cleverly taken
advantage of these nonidealities \cite{MCMC_memristor}\cite{clone} to create analog computing devices
beyond VMM accelerators. 

Stochastic differential equations (SDEs) are used in
many areas of applied mathematics, modeling systems with inherent
stochastic terms or systems in the presence of noise \cite{sdes_oksendal}. Their wide use
is despite the fact that there are no analytic solutions for most
models of interest, meaning that numerical methods are
necessary. Monte Carlo methods are one method of tackling the problem,
where many realizations of the process are simulated, and then
quantities of interest are computed from the resulting distributions.

However, Monte Carlo simulation suffers from two bottleneck
operations: solution of nonlinear equations for certain types of
solvers and generation of gaussian random variables \cite{sde_lab}. Therefore,
memristor crossbars can be used to solve the latter problem, where
gaussian random numbers can be generated directly in memory, reducing
time and energy cost compared to conventional algorithms implemented
on CMOS such as the Ziggurat algorithm \cite{ziggurat}\cite{ziggurat_hardware} or the Box-Muller transform. 

Previous works on hardware acceleration of Monte Carlo methods for SDEs have been focused on FPGAs \cite{fpga_sde_1} or fully analog computation \cite{columbia}\cite{columbia_sde}, but to date no work has been written evaluating the potential in this space for memristor crossbars. There have been tangentially related works on MCMC algorithms on in-memory computing platforms, such as \cite{MCMC_sram} (on SRAM) and \cite{MCMC_memristor} (on memristor crossbars), which are not directly applicable but provide motivation for this work. 

In this paper we examine generation of random numbers on an in memory
computing device, allowing the acceleration of Monte Carlo simulation of stochastic
differential equations. The inherent noisiness of crossbar
operations is shown to be not a barrier to their use as accelerators,
but an advantage which allows even more efficient calculation of
Monte Carlo simulations. 

\section{Background}

In this section we describe the characteristics of memristor crossbars
as well as basic mathematical results about stochastic differential
equations and numerical integration techniques. 

\subsection{Stochasticity in Memristor Crossbars}
Memristor crossbars are a type of resistive RAM that use programmable
resistance grids to perform highly efficient vector-matrix
multiplication. Because fabrication techniques are still in
development for such devices, the resolution of analog memristive
devices is quite low compared to conventional computers. The low
``resolution'' is due to limited programming tuning accuracy, with $99$\% of memristors within $4$\% of the target conductance in the most advanced
crossbars \cite{passive_crossbar}.

\begin{figure*}[h]
    \centering
    \includegraphics[width=\textwidth]{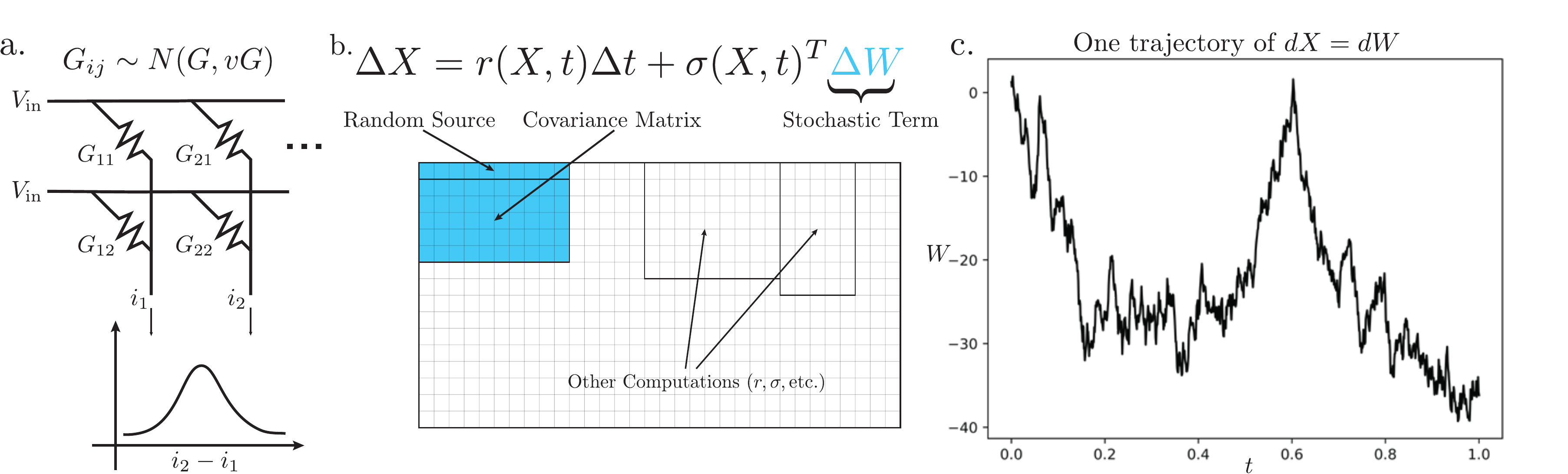}
    \caption{(a) The conductance of a memristor in its low conductance state is normally
      distributed with mean equal to the target conductance and
      standard deviation equal to some percentage $v$ of the target
      conductance. This can be exploited to create a gaussian random
      variable from the difference of the currents in the bit lines
      (in the figure $i_2 - i_1$). The difference of the currents is
      used to encode negative numbers, and makes summation with non
      random variables below simpler.(b) An example mapping of a random vector block. The random source memristors are reprogrammed to generate random vectors, and the outputs are multiplied by a covariance matrix to correct differences in variance. (c) A Wiener process, the solution to the simplest SDE $dX = dW$. The samples dW are taken from a pair of memristors on a crossbar, which are added up over time to generate the full trajectory. The trajectory was created by taking $1000$ time steps with value distributed as N(0,dt), which could be drawn from the conductance distribution of the memristors. }
    \label{fig:my_label}
\end{figure*} 

In OxRAM devices, the stochasticity in the cycle-to-cycle variability can be explained by variations in the conductive filaments. When the device is set and reset, the size and location of the filament changes, causing a change in the conductance state. Because of this random property, the high conductance state (HCS) displays a gaussian distribution \cite{hf_memristor}, with standard deviation equal to anywhere from 10\% to 40\% of the mean, which can be controlled via the write current \cite{MCMC_memristor}.

Although read operations can cause drifts in neighboring memristors in the same rows and columns, we neglect this effect because it is tolerable for a single read operation\cite{drift} even at larger ($>$1V) voltages that use up the random sample drawn from the memristor conductance distribution. 

The crossbar model used in this paper was written in PyTorch \cite{pytorch}. It supports variable input, output, and line resistance. The programming is modeled as linear with finite precision enforced by perturbing the write conductance by some percentage error, in this work a pessimistic 5\%. It should be noted that this variation is distinct from the cycle to cycle variability described above, which occurs in a region where the IV curve is not linear and accurate programming is very difficult. 

The read step is modeled using a 1-bit DAC, i.e. floating point vector inputs are converted into n-bit fixed precision binary inputs, and the n outputs are scaled and added together off the crossbar \cite{ISAAC}. All simulations in this paper used a line resistance of $5$ $\Omega$, an input and output resistance of $1$ K$\Omega$\cite{crossbarparams}, unless stated otherwise. 

\subsection{Stochastic Differential Equations}
Stochastic differential equations are a kind of differential equation
which involves a random term, used to model stochastic processes such
as ion drift, financial markets, and neuronal activity. The Itô
stochastic differential equation is

\begin{equation}
    dX = r(t, X)dt + \sigma(t, X)^{T} dW,
    \label{eqn:diffeq}
  \end{equation}
  
\noindent
where $r$ and $\sigma$ are vector valued functions and $dW$ is
the derivative of Brownian motion. The
time evolution of the probability distribution $p$ of $X$
follows the 2nd-order Fokker-Planck partial differential equation

\begin{equation}
    \frac{\partial p}{\partial t} = -\frac{\partial}{\partial x} (r p) + \frac{1}{2} \frac{\partial^{2}}{\partial x^2} (\sigma^{2} p)
    \label{eqn:kolgomorov}
  \end{equation}
which can be used to find analytic solutions to certain stochastic
differential equations.

\subsection{SDE Integrators}
Only a small class of SDEs can be solved analytically. Just as with
their deterministic counterparts, numerical methods must be used for
all but the simplest SDEs. The one step approximation to the
differential equation (\ref{eqn:diffeq}) yields the Euler-Maruyama
method:

\begin{equation}
    X_{n} - X_{n+1} = r(t_n, X_n) \Delta t + \sigma(t_n, X_n) \Delta W
\end{equation}
which is the simplest numerical stochastic integration method. $\Delta W$ is a normal random vector with zero mean and standard deviation $\Delta t$. The
method has an order of convergence of $O(\Delta t^{1/2})$ where
$\Delta t$ is the time step size.

\section{Solution}
In this section we describe the implementation of a stochastic
integrator over the crossbar, including generation and
characterization of noise, as well as the solver itself. A memristor
crossbar simulator is used, including input/output and line
resistance, tuning variability, quantized DAC inputs, and stuck on and stuck off
nonidealities. The crossbar is modeled as a semi-passive array, with 
$8\times8$ tiles of memristors.

\subsection{Random Sampling}
\begin{figure}[h]
    \centering
    \includegraphics[width=0.8\columnwidth]{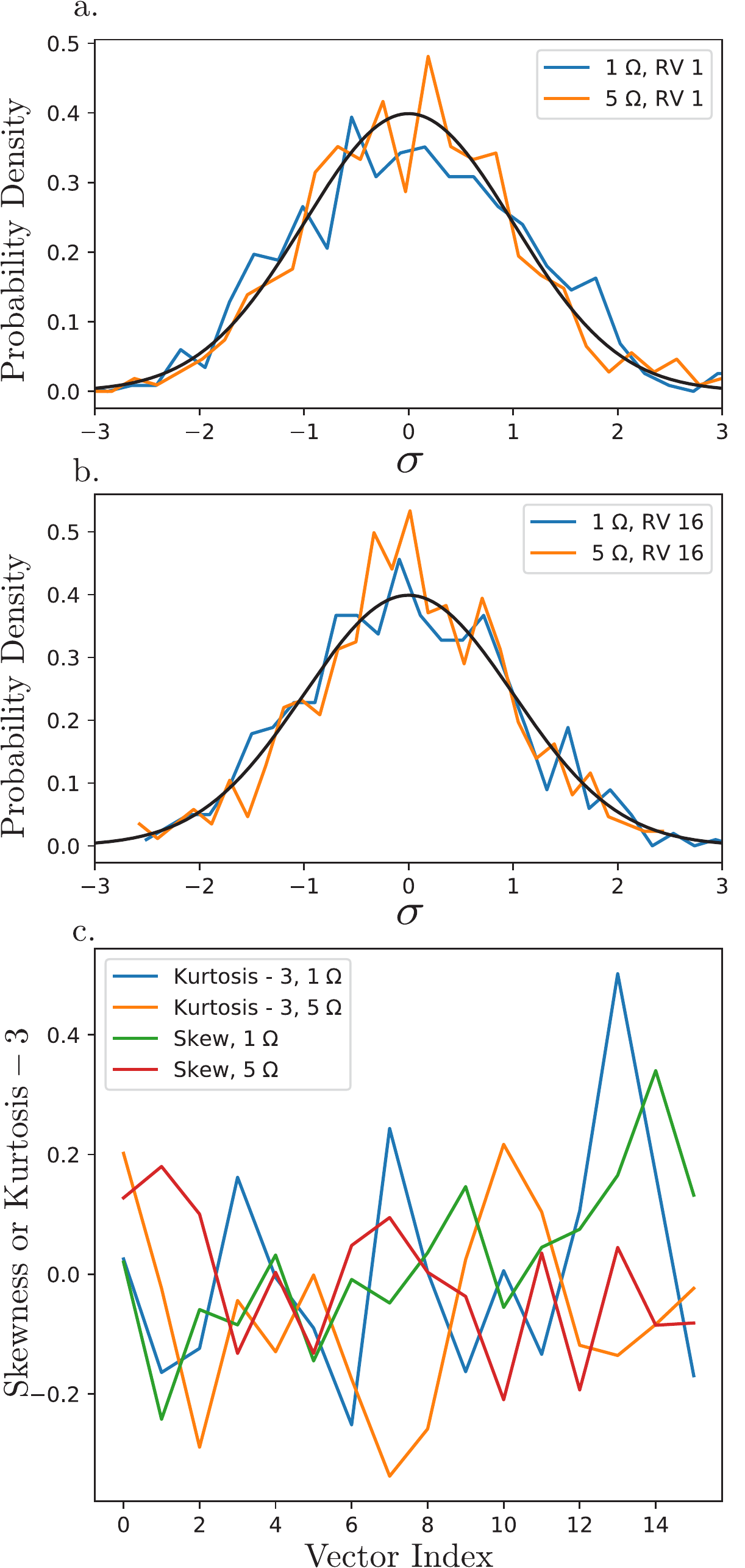}
    \caption{A 32x32 simulated crossbar was used to generate $500$ 16-length random gaussian vectors using the first word line. (a) Shows the change in the distribution due to line resistance in the first pair of memristors on the top word line. (b) Shows the same for the last pair of memristor. The mean and standard deviation are not changed by the increase in line resistance. (c) Shows the skew and kurtosis of the data in the top two plots. The kurtosis has 3 subtracted since the kurtosis of the unit normal distribution is 3. These higher order moments show no correlation with the position on the word line, indicating that line resistance does not negatively affect the properties of the gaussian sources. For reference, the standard deviation of the sample distribution of the kurtosis$- 3$ for 500 normally distributed variables is approximately 0.2, although the distribution is not symmetric. Therefore, most of the samples fell within one standard deviation of the of the sample mean of their test statistic.}
    \label{fig:r_variable}

\end{figure}

The integrator described above needs a source of gaussian random vectors. The
crossbar can provide this through the uncertainty in the low conductance state of a memristor, which follows a gaussian distribution centered around the target conductance with variability proportional to the conductance. The variability is a characteristic of the memristor and the programming current. A line of memristors can be programmed to the same conductance in order to generate a gaussian random vector with a certain mean and standard deviation, which can be rescaled to a unit normal vector. Although line resistance can cause voltage degradation in larger passive arrays, we do not observe this to have an effect on the distributions of the outputs, shown in Figure \ref{fig:r_variable}. The pair of memristors at the end of the word-line produce the same output variance, regardless of the line resistance.

For generation of gaussian random vectors with arbitrary covariance, the unit normal random vector can be multiplied by the lower triangular matrix of the
Cholesky decomposition of the covariance matrix. This matrix
multiplication has limited precision, resulting in a gaussian
distribution close to but not exactly equal to the desired gaussian
spectra. This is useful for modeling correlated processes.
\subsection{Function Evaluation}
The crossbar is not capable of computing general nonlinear functions
such as $r(t, X)$ and $g(t, X)$. Therefore, they must be evaluated on
a normal computer, using a computer-in-the-loop architecture. Linear
functions can be evaluated as usual on the crossbar, and the allocation of computational tasks to the crossbar should be decided based on the problem. Recent work on neural networks in the framework of stochastic differential equations are promising for memristor crossbar acceleration because they involve many linear weight-stationary operations \cite{neural_sde}. 

\section{Applications}

\subsection{Solving the Black-Scholes SDE}
The Black-Scholes SDE (Eq \ref{eqn:bs}) for geometric brownian motion is a classic stochastic differential equation \cite{finance}. The case we consider is for a market with a fixed interest rate $r$ and volatility $\sigma$, since this simple case has a closed form solution. 
\begin{equation}
  dX = r X dt - \sigma X dW \textrm{  subject to  } X(t=0) = X_0
  \label{eqn:bs}
\end{equation}

Because it has a closed form solution (Eqn. \ref{eqn:bs_sol}), we can use it to test
the accuracy of our solver by comparing the distribution with the
distribution of the final values of the trajectories simulated on the
crossbar.

\begin{equation}
  X = X_0 \exp(\sigma W + (r - \sigma^2 / 2) t )
  \label{eqn:bs_sol}
  \end{equation}

The parameters $r$ and $\sigma$ were mapped to the crossbar, and a 3rd pair of memristors set to their high resistance state (100 k$\Omega$) was used as the random source. The variability of the HRS was set at 25\%. Because the problem is 1-D, no covariance matrix is needed. Figure 3(a) shows the layout of the parameters on the crossbar. All unused memristors were set to their low resistance state (10 k$\Omega$), with 10\% standard deviation.

\begin{figure}[h]
    \centering
    \includegraphics[width=0.85\columnwidth]{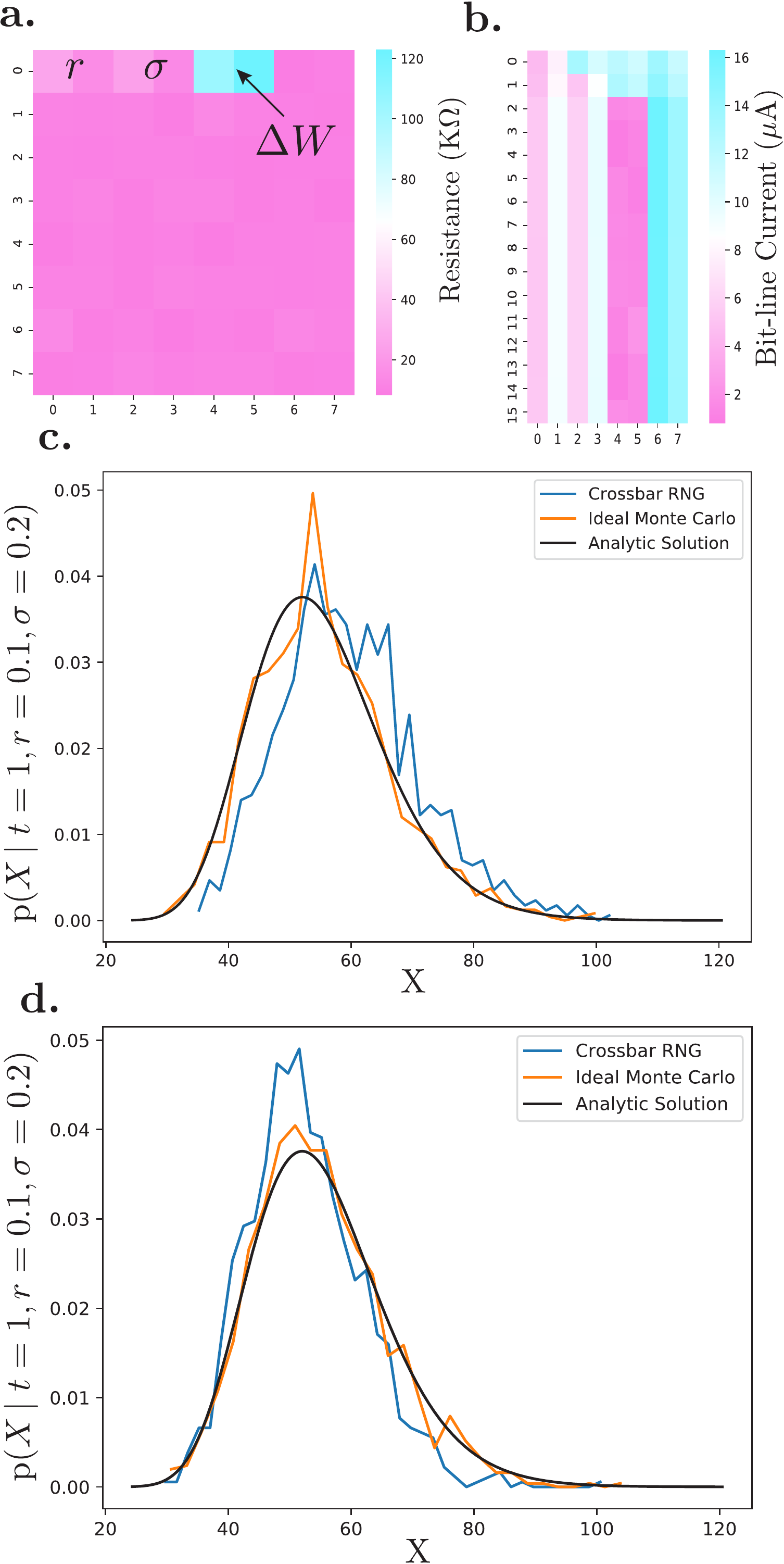}
    \caption{Simulation of the Black-Scholes equation. The equation
      was simulated from t=0 to t=1, with 100 time steps. The
      parameters were r=0.1, $\sigma$=$0.2$. (a) Heatmap of resistances of the devicess in a simulated 8x8 array, showing the location of $r$, $\sigma$, and the random source for $\Delta W$. (b) Shows the first 16 bitline currents for read operations. The first two operations are single read operations used to model the energy consumption during writing of $r$ and $\sigma$. The subsequent read operations are from the random source, and the stochasticity in the output currents can be observed. In
      (c), the
    blue line shows the distribution of the final positions of the
    trajectories using the crossbar random number generator and solver
    weights. The black line
      shows the analytic solution given that W is brownian motion. In (d) only the random numbers were generated
    on the crossbar. Note the slight skew in the top plot. The orange line shows the
  final positions of the trajectories calculated using fully
  digital calculations. }
    \label{fig:mc_bs}
\end{figure}

The results of Monte Carlo simulation using the Euler-Maruyama method
are shown in Figure \ref{fig:mc_bs}. For calculation of $rX$ and $\sigma X$, the input was encoded in 16 bits. The simulation involving only generation of random numbers on the crossbar is in good agreement with the analytic solution. However, the simulation involving computation of $r$ and $\sigma$ displays a slight skew. This is due to the crossbar model perturbing the conductances to model imperfect writing. When this parameter is turned down the skew disappears. Although the variability is small, over $100$ time steps the effect on the distribution can be significant. Because these values are so important, more care is required in programming them so that they land closer to their target conductance. 

\subsection{Static Power Consumption}
Gaussian random sampling is expensive on digital devices because
several calculations are needed to transform from a uniform
distribution for which pseudo-random number generators exist to normal
random variables. Because of this, the implicit gaussian distribution
on the crossbar allows for some savings in computation despite
requiring a write operation each time a new sample is needed. 

The static power consumption of the memristor array was calculated following the model for the eVM synaptic array in NeuroSim \cite{NeuroSim}, although their software was not used. The memristors were modeled as taking $100$ of $1$ V write pulses of $200$ ns to get from low to high resistance state, as well as a $1$ ms $0.2$ V read operation between each write pulse. This is a pessimistic estimate because writing to the HRS does not require so many read operations at first because it is unlikely that the resistance will overshoot the HRS. Unselected resistors were set at the low resistance state. 

For the 2-memristor random variable shown here, this resulted in an average writing energy of 0.8 $\mu$J, which was dominated by the energy consumed by the read operations. The full simulation of the Black-Scholes equation involved roughly $200,000$ write operations, resulting in an energy of consumption 0.16J. Because all unused memristors were set at the low resistance state, the power consumption is a worst-case scenario. The energy consumed by read operations was 3$\mu$J. Scaling to higher dimensional problems would increase this figure linearly.

\section{Conclusion}
In this paper, simulation of a simple SDE was shown over memristor hardware, 
where the white noise was drawn from the intrinsic uncertainty in programming. The effects of crossbar nonidealities such as line resistance were examined in order to determine their effects on the shape of the gaussian distribution. Despite nonidealities associated with inaccurate correction for variance, the solver produces a good fit to the analytic solution. This paper demonstrates that programming variability can be taken advantage of for solutions of SDEs while still being useful for vector matrix multiplication operations.

There are several directions this work can be taken in the future. Integration of a more complete crossbar model including 1T1R configurations and stochastic memristor models would improve the characterization of distributions resulting from crossbar operations. Although not discussed in this paper, the methods used here could be applied to higher dimensional SDEs where no analytical solutions exist, where the VMM acceleration of the crossbar would be significant.

%\section*{Acknowledgment}
%will clean up bibliography later

% that's all folks
\end{document}